\documentclass[conference]{IEEEtran}

\usepackage[english]{babel}
\usepackage[rflt]{floatflt}
\usepackage{graphicx}
\usepackage{url}
\usepackage{multirow}
\usepackage[rflt]{floatflt}
\usepackage{graphicx}
\usepackage{multirow}
\usepackage[table]{xcolor}
\usepackage{supertabular}
\usepackage{spverbatim}
\usepackage{paralist}
\usepackage{multicol}
\usepackage{csquotes}
\usepackage{flushend}
\usepackage{cite}

\usepackage{enumitem}

\title{An Efficient, Secure and Trusted Channel Protocol for Avionics Wireless Networks}

\begin{document}

\author{\IEEEauthorblockN{Raja Naeem Akram\IEEEauthorrefmark{2}, Konstantinos Markantonakis\IEEEauthorrefmark{2}, Keith Mayes\IEEEauthorrefmark{2}\\
Pierre-Fran{\c c}ois Bonnefoi\IEEEauthorrefmark{3}, Damien Sauveron\IEEEauthorrefmark{3}\IEEEauthorrefmark{4} and Serge Chaumette\IEEEauthorrefmark{4}
}
\IEEEauthorblockA{\IEEEauthorrefmark{2}Information Security Group Smart Card Centre, Royal Holloway, University of London, Egham, United Kingdom\\
\IEEEauthorrefmark{3}XLIM (UMR CNRS 7252 / Universit\'e de Limoges), D\'epartement Math\'ematiques Informatique. Limoges, France\\
\IEEEauthorrefmark{4}LaBRI (UMR CNRS 5800 / Universit\'e de Bordeaux), Talence, France \\
Email: \{r.n.akram, k.markantonakis, keith.mayes\}@rhul.ac.uk,\\ \{pierre-francois.bonnefoi, damien.sauveron\}@unilim.fr,  serge.chaumette@labri.fr}}

\maketitle

\begin{abstract}

Avionics networks rely on a set of stringent reliability and safety requirements. In existing deployments, most of these networks are
based on a wired technology, which supports these requirements. Furthermore, this technology simplifies the security management of the network since certain assumptions can be safely made, including the inability of an attacker to access the network, and the fact that it is almost impossible for an attacker to introduce a node into the network. The proposal for Avionics Wireless Networks (AWNs, currently under consideration by multiple aerospace working groups, promises a reduction in the complexity of electrical wiring harness design and fabrication, a reduction in the total weight of wires, increased customization possibilities, and the capacity to monitor otherwise inaccessible moving or rotating aircraft parts such as landing gear and some sections of the aircraft engines. While providing these benefits, the AWN must ensure that it provides levels of safety that are at minimum equivalent to those offered by the wired equivalent. In this paper, we propose a secure and trusted channel protocol that satisfies the stated security and
operational requirements for an AWN protocol. There are three main objectives for this protocol. First, the protocol has to provide the assurance that all communicating entities can trust each other, and can trust their internal (secure) software and hardware states. Second, the protocol has to establish a fair key exchange between all communicating entities so as to provide a secure channel. Finally, the third objective is to be efficient for both the initial start-up of the network and when resuming a session after a cold and/or warm restart of a node. The proposed protocol is implemented within a demo AWN, and performance measurements are presented based on this implementation. In addition, we formally verify our proposed protocol using CasperFDR.
\end{abstract}

\section{Introduction}
A modern aircraft can be considered as a highly reliable and mission-critical digital network in the air. The Aircraft Data Network (ADN) interconnects different aircraft sub-systems, including flight control, the crew network and the passenger entertainment network. In recent years investigations into the feasibility of moving some non-critical networks from wired technology to wireless-based technology have been carried out. Such a network is referred to as an Avionics Wireless Network (AWN), which is the main focus of this paper.

Whatever the network deployment topology and the communication technology that are used, one element is common: the physical wire that connects two or more avionics sub-systems. Wiring an aircraft can be costly in that it includes wiring harness designs, cable fabrication and the associated cost of additional weight. Furthermore, to provide dual redundancy, these wires have to connect any two devices by means of two physically separate paths in the aircraft. Wires and related connectors potentially represent 2-5 percent of an aircraft's weight \cite{ITU2010}. As the wiring of an aircraft is a time- and labor-intensive activity, post-deployment upgrades or installation of new wire routes or new avionics sub-systems may be costly \cite{Dang2012}. As reported by \cite{ITU2010}, roughly 30 percent of wires are potential candidates for wireless substitutes. Therefore, as highlighted in \cite{RNAkram2015}, wireless solutions have more than reasonable prospects as long as security, safety and high reliability can be maintained.  

Whether an ADN or an AWN is used, the main objective is to communicate data between aircraft sub-systems in a secure, reliable and efficient manner. Going wireless brings its own set of unique challenges, among which a major one is to ensure the confidentiality and integrity of communications; any attacker within wireless range of the AWN can easily eavesdrop and/or (potentially) modify the exchanged information. To protect against such an attack, we require a strong, efficient and trustworthy mechanism to establish secure links between the communicating nodes in an AWN. Secure channel protocols can be used for this purpose, and in this paper we propose such a protocol for AWN environments. In this paper, we are not going to discuss the wireless jamming attacks. Although they are a valid threat but they do not directly attack the confidentiality and integrity of communication channel - wireless jamming attack is a thread to channel availability. For this reason they are beyond the scope of this paper. 

\subsection{Contribution}
In this paper, our main goals are to propose a secure and trusted channel protocol for AWNs, and to compare its security and performance with several other existing protocols.

The salient contributions of this paper are as follows:

\begin{enumerate}
	\item proposing a Secure and Trusted Channel Protocol (STCP) that along with establishing a secure channel between the communicating entities (end-points) also provides security assurance that each end-point is secure and trusted;
	\item defining comparison criteria for secure channel protocols along with the related security and performance analysis;
	\item validating the proposed protocol with a formal tool, CasperFDR and producing an implementation in a real AWN to enable measurements to be obtained.
\end{enumerate}

\subsection{Structure of the Paper}
Section~\ref{sec:Related_Work} briefly presents the rationale behind this paper and the existing work carried out in the avionics industry (in the context of AWNs) and secure channel protocols from a traditional computer security perspective. In section \ref{sec:Trusting_a_Device}, we look into how a Trusted Platform Module (TPM) can provide a trusted boot that is then used to assure communication partners that the device is secure and trustworthy. Section \ref{sec:Secure_and_Trusted_Channel_Protocol} discussed the security comparison criteria and then the proposed protocol. In section \ref{sec:Protocol_Evaluation}, we first analyze the proposed protocol informally, than formally using CasperFDR and we compare it with different protocols based on the security comparison criteria previously defined. Finally in section \ref{sec:Conclusion} we present future research directions and conclude the paper.

\section{Rationale and Related Work}
\label{sec:Related_Work}
In this section, we discuss the rationale behind the proposed protocol and review the existing work in two different areas: AWNs and Secure Channel Protocols (SCPs). 

\subsection{Rationale}
\label{sec:Rationale}
A Secure Channel Protocol (SCP) by definition provides either or both of entity authentication and key exchange between communicating parties (end points). An  SCP preserves the confidentiality and integrity of the messages on the considered channel but not at the end points. 

Nevertheless, there can be implicit assurance in the integrity and security of the end points as described by ETSI TS 102 412 \cite{ETSITS102412} in the domain of the smart card industry. This document states that the smart card is a secure end point under the assumption that it is a tamper-resistant device. This type of assurance can be extrapolated to other devices that are implicitly trusted because of offline business relationships or because of a property of the device itself. 

However, for a critical system like avionics it is not just implicit trust that should be required but also explicit trust validation,  to counter any potential threat. The explicit trust assurance should be provided by the (aircraft) device that is participating in the AWN communication. This would build in an assurance that only secure and trusted devices (explicitly trusted devices with per-protocol run assurance) will participate in the AWN, potentially countering physically altered devices and/or re-introduction of a decommissioned device as discussed in \cite{RNAkram2015,RNAkram2016a}.  

In contrast, in the ADN, the assumption of implicit assurance might be valid.
However, for a robust security and reliability mechanism an explicit security assurance mechanism should be considered. 

A trusted channel is a secure channel that is cryptographically bounded to the current state of the communicating parties \cite{Gasmi2007}. This state can be a hardware and/or a software configuration, and ideally it requires a trustworthy component to validate it is effectively as claimed. Such a component, in most instances, is a TPM \cite{TPMSpec2011} as demonstrated in \cite{10.1109/CMC.2010.232,Armknecht2008,Akram2012b}

In an AWN, individual devices will have prior relationships with each other: in the avionics industry any system deployment is stringently controlled, regulated and protected. Therefore, assuming that one single trusted entity would deploy the AWN environment is as per the avionics industry's practice. However, when establishing a secure channel, individual devices should still ensure that they are not only communicating with an authenticated device but also that the current state of this device is secure.

\subsection{Related Work on AWN Security Concerns}
\label{sec:Related Work on Security Concerns}
Security and trust have been subject to some analysis by both the academic community and the industry. A brief overview of aircraft information security and some improvements were proposed in \cite{Olive2006}. Security assurance research from airplane production to airplane operation was presented in \cite{Lintelman2006,ladstaetter2011security}. A general discussion of the security issues related to the aircraft network and aircrafts' connectivity with the Internet is provided in \cite{thanthry2006security}, while \cite{Thanthry2004,Robinson2007a} discusses the impact of WSNs (Wireless Sensor Networks) and related security concerns in aircraft. Security and safety are intrinsically linked to each other in general and specifically in the context of the aviation industry \cite{brostoff2001safe,Pfitzmann2004,Paulitsch2012}. The application and impact of cryptography, especially public key cryptography for avionics networks, was evaluated in \cite{Robinson2007}.

The management of security and the general deployment of AWNs based on wireless-as-a-comm-link have been analyzed in \cite{RNAkram2015}, which discusses the security and trust challenges faced by AWNs. In addition, a crucial component that supports aircraft devicesâ security is the trusted boot process discussed in \cite{RNAkram2016a}. The security, trust and assurance issues related to the fact of bringing a user device into an aircraft network are evaluated in \cite{RNAkram2016b}.% However, general guidelines and experience in designing wired aircraft networks and WSN are still worth mentioning, as current and related future work is based on them.\todo[inline]{DS:Do we keep the previous sentence?}

\subsection{Related Work on Secure Channel Protocols}
\label{sec:RelatedWorkonSecureChannelProtocols}
In this section, we restrict the discussion to the protocols that are proposed for general-purpose computing environments or to those that are used as points of comparison in the discussions to come. 

The concept of trusted channel protocols was proposed by Gasmi et al. \cite{Gasmi2007} along with the adaptation of the TLS protocol \cite{SSLTSLRFC2008}. Later Armknecht et al. \cite{Armknecht2008} proposed another adaptation of OpenSSL to accommodate the concept of trusted channels; similarly, Zhou and Zhang \cite{10.1109/CMC.2010.232} also proposed an SSL-based trusted channel protocol.

In section \ref{sec:Revisiting_the_Requirements_and_Goals}, we will compare the proposed STCP with the existing protocols. These protocols include the Station-to-Station (STS) protocol \cite{DiffiAAAKE92_209}, the Aziz-Diffie (AD) protocol \cite{Whitfield94privacyand}, the ASPeCT protocol \cite{Horn1998}, Just-Fast-Keying (JFK) \cite{Aiello:2004:JFK:996943.996946}, trusted TLS (T2LS) \cite{Gasmi2007}, GlobalPlatform SCP81 \cite{GPCSPE011}, the Markantonakis-Mayes (MM) protocol \cite{kostas2004}, and the Sirett-Mayes (SM) protocol \cite{Sirett2006}.

This selection of protocols is intentionally broad so as to include well-established protocols like STS, AD and JFK. We also include the ASPeCT protocol, which is designed specifically for mobile networks' value-added services. Similar to our proposal where we require trust assurance during the protocol run, T2LS meets this as it provides trust assurance, whereas other protocols like SCP81, SM, and MM are specific to smart cards and are representative embedded low-power devices.  In addition, we have included the secure and trusted channel protocol, P-STCP \cite{Akram2012b}, which is designed for resource-restricted and security-sensitive environments, and has some similar design requirements to those of the proposed protocol.

\section{Trusting a Device (Trusted Boot)}
\label{sec:Trusting_a_Device}
In this section, we discuss how a TPM provides a secure boot process and how it provides assurance to external entities that the device is secure and trustworthy. 

\subsection{Trusted Platform Module}
\label{sec:Trusted_Platform_Module}
The TPM is a trusted, reliable and tamper-resistant component that can provide trustworthy evidence of the state of a given system on which it is present. The interpretation of this evidence is neither controlled nor dictated by the TPM but by the entity receiving and thus assessing it. Trust in this context can be defined as an expectation that the state of a system is as it is supposed to be, i.e. secure. Therefore, in a very simplistic sense a TPM is a trustworthy reporting agent (witness), not an evaluator or an enforcer of security policies. In the field of trusted computing, this is referred to as providing a root of trust on which an inquisitor relies to validate the current state of a system.

For in-depth discussion of the architecture of TPMs and their functionality please refer to \cite{TPMSpec2011}. In this paper, we focus on the secure boot process as it is carried out by the TPM  and as discussed in the subsequent section. 

\subsection{Secure Boot (TPM Integrity Measurement Operation)}
When a  device with a TPM boots up, the first component to power up is the system BIOS (Basic Input/output System). On a trusted platform (a platform that contains a TPM), the boot sequence is initiated by the Core BIOS (\emph{i.e.} CRTM: Core Root of Trust Measurement), which first measures its own integrity. This measurement is stored in PCR$_0$\footnote{A Platform Configuration Register (PCR) is a 160-bit (20 bytes) data element that stores the result of the integrity measurement, which is a generated hash of a given component (\emph{e.g.} the BIOS, the operating system, or an application). A group of PCRs form the integrity matrix. The process of extending PCR values is as follows: $PCR_{i}$ = $Hash({PCR^{'}}_{i} || X)$, where $i$ is the PCR index, ${PCR^{'}}_{i}$ represents the old value stored at index $i$, and $X$ is the sequence to be included in the PCR value. ``$||$'' indicates the concatenation of two data elements in the given order. The starting value of all PCRs is zero.} and it is later extended to include the integrity measurement of the rest of the BIOS. The Core BIOS then measures the circuit-board's (motherboard) configuration setting\footnote{To measure that correct hardware configuration was present at boot time.}, and this value is stored in PCR$_1$. After these measurements, the Core BIOS loads the rest of the code of the BIOS.

\begin{figure}
%\vspace{-18pt}
	\centering
		\includegraphics[width=0.75\columnwidth]{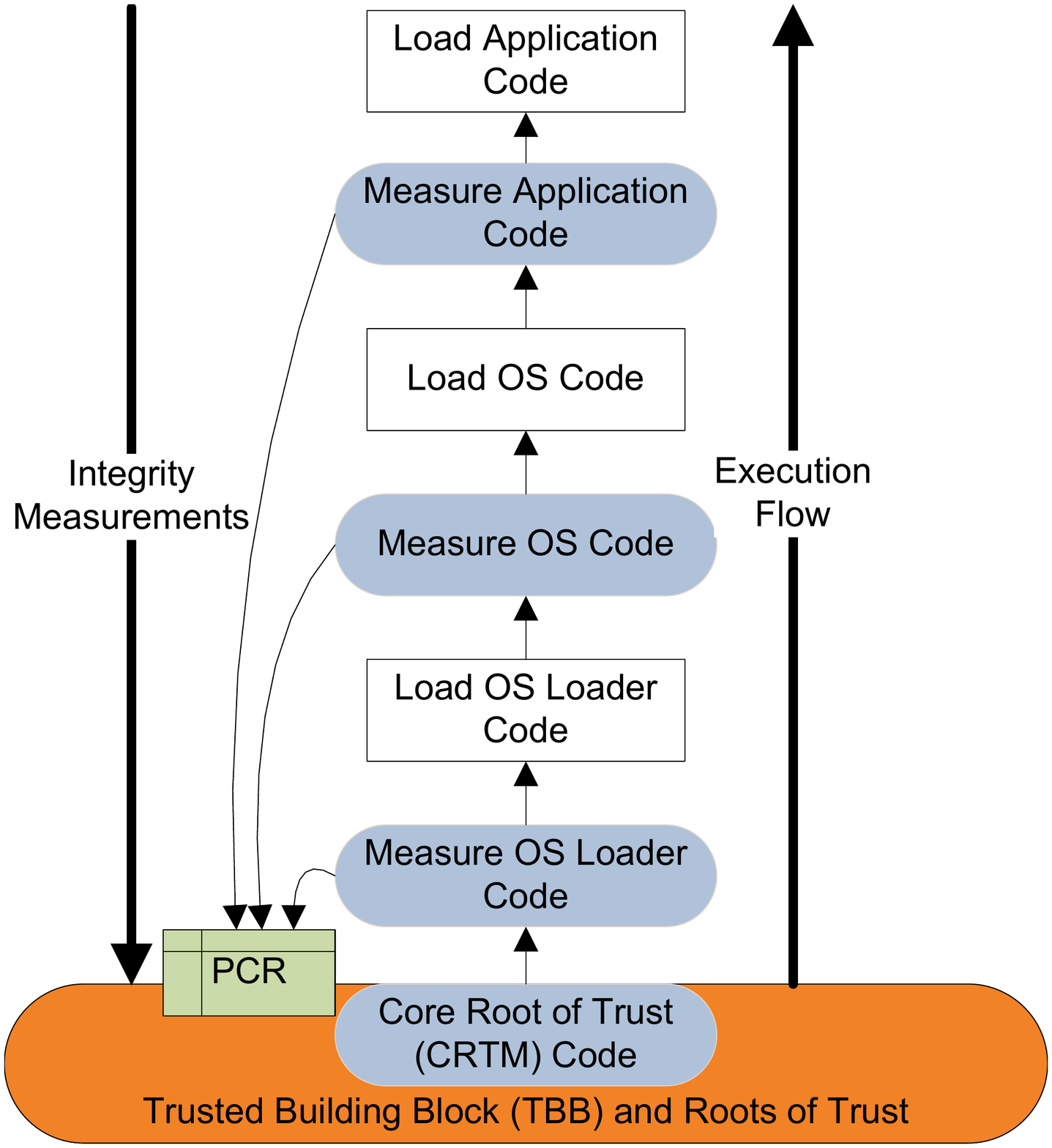}
	\caption{Trusted Platform Boot Sequence (figure from \cite{akram2014introduction})}
	\label{fig:BootLoading}
	\vspace{-14pt}
\end{figure} 

The BIOS will subsequently measure the integrity of the ROM firmware and of the ROM firmware configuration, storing them in PCR$_2$ and PCR$_3$ respectively. At this stage, the base configuration of a device is established and the CRTM will proceed with integrity measurement and loading of the Operating System (OS).

The CRTM measures the integrity of the ``OS Loader Code'', also termed the Initial Program Loader (IPL), and stores the measurement in the relevant PCR.
The designated PCR index is left to the discretion of the OS developers. Subsequently, the device will execute the ``OS Loader Code'' and if successful, the TPM will measure the integrity of the ``OS Code''. After this measurement is made and stored, the ``OS Code'' executes. Finally, the relevant software that initiates its execution will first be subjected to an integrity measurement, and the resulting value will be stored in a PCR and then the software will be allowed
to execute. This process is shown in Figure \ref{fig:BootLoading}, which illustrates the execution flow and the storage of the integrity measurements.

By creating a chain of integrity measurements, a TPM provides a trusted and reliable view of the current state of the system. Any piece of software, whether part of the OS or an application, has an integrity measurement stored in a PCR at a particular index.

As discussed above, a TPM does not make any decisions: it only measures, stores, and reports integrity measurements in a secure and reliable manner. When a TPM reports an integrity measurement, it is recommended that it generates a signature on the value, thus avoiding replay and man-in-the-middle attacks \cite{TPMSpec2011}. The process by which an inquisitor can request a device attestation and how a TPM provides this evidence is discussed in the next section.

\subsubsection*{Reporting and Attestation Operations}
\label{sec:Attestation}
The attestation process, whether initiated by the relevant external entity (including human users or other devices) locally or remotely, involves the generation of a signature by the TPM using the Attestation Identification Key (AIK) of the (associated/requested) PCR values \cite{akram2014introduction}. The signature assures the requesting entity of the validity of the integrity measurement stored in the PCRs. The choice of the AIK and PCR index is dependent on the device, OS or application developer.

The signature key and PCR values are stored in a tamper-resistant memory inside the TPM. Therefore, an attacker would have to circumvent the tamper-resistant property of the TPM to impact the outcome of this attestation process.

\section{Secure and Trusted Channel Protocol}
\label{sec:Secure_and_Trusted_Channel_Protocol}
In this section, we begin the discussion with the security comparison criteria, followed by the protocol notation, pre-setup and then the actual protocol proposal. This section concludes with a discussion of how the secure channel is re-established if one of the devices is restarted or resets the protocol.   

\subsection{Security Comparison Criteria}
\label{sec:Comparison_Criteria}

For a protocol to support the AWN framework, it should meet, at minimum, the security and operational requirements listed below:

\begin{enumerate}[label=\textbf{G\arabic*})]

\item {\bf {\itshape Mutual Entity Authentication}:} All nodes in the network should be able to authenticate to each other to avoid masquerading by a malicious entity.

\item {\bf Asymmetric Architecture:} Exchange of certified public keys between the entities to facilitate the key generation and entity authentication process. 

\item {\bf Mutual Key Agreement:} Communicating parties will agree on the generation of a key during the protocol run. 

\item {\bf Joint {\itshape Key Control}:} Communicating parties will mutually control the generation of new keys to avoid one party choosing weak keys or predetermining any portion of the session key.  

\item {\bf {\itshape Key Freshness}:} The generated key will be fresh to the protocol session to protect against replay attacks.

\item {\bf Mutual {\itshape Key Confirmation}:} Communicating parties will provide implicit or explicit confirmation that they have generated the same keys during a protocol run. 

\item {\bf Known-Key Security:} If a malicious user is able to obtain the session key of a particular protocol run, it should not enable him to retrieve long-term secrets ({\itshape private keys}) or {\itshape session keys} (future and past).

\item {\bf Unknown {\itshape Key} Share Resilience:} In the event of an unknown key share attack, an entity $\mathcal{X}$ believes that it has shared a key with $\mathcal{Y}$, where the entity $\mathcal{Y}$ mistakenly believes that it has shared the key with entity $\mathcal{Z} \neq \mathcal{X}$. Proposed protocols should adequately protect against this attack.

\item {\bf {\itshape Key} Compromise Impersonation (KCI) Resilience:} If a malicious user retrieves the long-term key of an entity $\mathcal{Y}$, it will enable him to impersonate $\mathcal{Y}$. Nevertheless, key compromise should not enable him to impersonate other entities to $\mathcal{Y}$ \cite{Blake-Wilson:1997:KAP:647993.742138}.

\item {\bf {\itshape Perfect Forward Secrecy}:} If the long-term keys of communicating entities are compromised, this will not enable a malicious user to compromise previously generated session keys. 

\item {\bf Mutual {\itshape Non-Repudiation}:} Communicating entities will not be able to deny that they have executed a protocol run with each other. 

\item {\bf Partial Chosen Key (PCK) Attack Resilience:} Protocols that claim to provide joint key control are susceptible to this type of attack \cite{Mitchell1998}. In this type of attack, if two entities provide separate values to the key generation function then one entity has to communicate its contribution value to the other. The second entity can then compute the value of its contribution in such a way that it can dictate its strength (i.e.\ it is able to generate a partially weak key). However, this attack depends upon the computational capabilities of the second entity. Therefore, proposed protocols should adequately prevent PCK attack.

\item {\bf Trust Assurance (Trustworthiness):} The communicating parties not only provide security and operation assurance but also validation proofs that are dynamically generated during the protocol execution.
 
\item {\bf Denial-of-Service (DoS) Prevention:} The protocol should not require the individual nodes to allocate a large set of resources to the extent that it might contribute to a DoS attack.

\item {\bf Privacy:} A third party should not be able to know the identities of the AWN nodes.

\end{enumerate}

For a formal definition of the terms (italicized) used in the above list, the reader is referred to \cite{Menezes1996}. The requirements listed above are later used as a point of reference to compare the selected protocols in Table \ref{tab:ProtocolComparisonOnTheBasiesOfStatedGoals}.

For the performance evaluation that we have conducted, the main measurements are related to the time required to establish a secure channel once the wireless link is established and they are discussed in section \ref{sec:Practical_Implementation}.

\begin{table*}[t]
\begin{center}
\caption{Secure and Trusted Channel Protocol (STCP).}
	\label{tab:STCP}
\resizebox{1\hsize}{!}{%
\begin{tabular}{lrcl}
\hline

1.&$AD1 \rightarrow AD2 $& : & $ AD1_{i}\| AD2_{i}\| N_{AD1}   \|   g^{r_{AD1}}  \| VR_{AD1-AD2}  \|  S_{Cookie} 	$\\
2.&$AD2 \rightarrow AD1$ &:& $ AD2_{i}\| AD1_{i}\|  N_{AD2}  \| g^{r_{AD2}} \|   [Sign_{AD2}(AD2-Data) \| Sign_{TPM_{AD2}}(AD2-Validation)]^{K_{e}}_{K_{a}} \| VR_{AD2-AD1} \|  S_{Cookie} $\\
  &                      &:& $ AD2-Data = H(AD2_{i} \|  AD1_{i}  \| g^{r_{AD1}} \| g^{r_{AD2}} \| N_{AD1} \| N_{AD2}$)\\
  &	 					 &:& $ AD2-Validation = SAS_{AD2-AD1} \| N_{AD1} \| N_{AD2} $\\
3.&$AD1\rightarrow AD2$ &:& $ [Sign_{AD1}(AD1-Data) \| Sign_{TPM_{AD1}}(AD1-Validation)]^{K_{e}}_{K_{a}}  \|S_{Cookie}	$\\
  &                      &:& $ AD1-Data = H(AD1_{i} \|  AD2_{i}  \| g^{r_{AD2}} \| g^{r_{AD1}} \| N_{AD2} \| N_{AD1}$)\\
  &	 					 &:& $ AD1-Validation = SAS_{AD1-AD2} \| N_{AD2} \| N_{AD1} $\\ 
								\hline 
\end{tabular}
}
\end{center}
\end{table*}

\subsection{Protocol Notation}
\label{sec:Protocol_Notation}
The notations used in the protocol description are listed in Table \ref{tab:NotationTable};

\begin{table}[h]

%\tablecaption{Protocol Notation}
\caption{Notation used in protocol description.}
	\label{tab:NotationTable}
		\centering
\begin{tabular}{p{1.3cm} p{0.01cm} p{6.1cm}}\hline
$AD1$	&:&Denotes an aircraft device '1'.\\
$AD2$	&:&Denotes an aircraft device '2'.\\
$A\rightarrow B$	&:&Message sent by an entity A to an entity B.\\
$TPM_X$ &:&Denotes a TPM of an entity $X$\\
$X_{i}$&:&Represents the identity of an entity $X$.\\  
$g^{r_X}$ &:&Diffie-Hellman exponential generated by an entity $X$.\\
%$C_{X}$	&:&Signature key certificate of an entity $X$.\\
$N_{X}$	&:&Random number generated by an entity $X$.\\ 
$X\|Y$	&:&Represents the concatenation of the data items X, Y in the given order.\\ 
$\left[ M \right] ^{K_{e}}_{K_{a}}$ &:& Message $M$ is encrypted by the session encryption key $K_{e}$ and then MAC is computed using the session MAC key $K_{a}$. Both keys $K_{e}$ and $K_{a}$ are generated during the protocol run. \\ 
$Sign_{X}(Z)$	&:&	Signature generated on data Z by the entity $X$ using a signature algorithm \cite{Furlani2009}. \\ 
$H(Z)$ &:& Is the result of generating a hash of data Z. \\ 
$H_{k}(Z)$	&:&	Result of generating a keyed hash of data Z using key $k$. \\ 
$S_{Cookie} $&:& Session cookie generated by one of the communication entities. It indicates the session information and facilitates protection against DoS attacks along with (possibly) providing the protocol session resumption facility. \\
$VR_{A-B}$&:& Validation request sent by entity A to entity B. In response entity B provides a security and reliability assurance to entity A.\\
$SAS_{A-B}$&:& Security assurance (PCR values) generation by entity A that  provides trust validation to the requesting entity B.\\
\hline
\end{tabular} 
\end{table}

\subsection{Pre-Protocol Setup}
\label{sec:Protocol_Requirements}
The proposed protocol requires certain pre-protocol setup operations as listed below:

\begin{enumerate}
\item Each aircraft device that is part of the AWN has a TPM. 
\item Each device in the AWN is pre-configured with the signature verification keys of its communication partners  (\emph{i.e.} public keys of other aircraft devices). 
\item Each device is also pre-configured with the signature verification keys of the TPMs of its communication partners (\emph{i.e.} the public key corresponding to the AIK key used to sign the PCR values stored in the TPM) along with their own trusted and secure PCR values (\emph{i.e.} the values for their trusted and secure state).
\end{enumerate}

\subsection{Proposed Protocol}
\label{sec:Proposed_Protocl}

The messages of the protocol are listed in Table \ref{tab:STCP} and described below.

\paragraph{\bf Message 1}
\label{sec:Message1}
The AD1 generates a random number $N_{AD1}$ and computes the Diffie-Hellman exponential $g^{r_{AD1}}$.  The ``$ H(g^{r_{AD1}}\| N_{AD1}\| AD1_{i} \| AD2_{i})$'' serves as a session cookie ``$S_{Cookie}$'', and it is appended to each subsequent message sent by both devices. It indicates the session information, facilitates protection against DoS attacks and (possibly) provides the protocol session resumption facility, which is required if a protocol run is interrupted before it successfully concludes. Finally, AD1 will request AD2 to provide assurance of its current state.

\paragraph{\bf Message 2}
\label{sec:Message2}
In response, AD2 generates a random number, and a Diffie-Hellman exponential $g^{r_{AD2}}$. It can then calculate the $k_{DH}=(g^{r_{AD1}})^{r_{AD2}} \ (mod\  n)$ which will be the shared secret from which the rest of the keys will be generated. The encryption key is generated as $K_{e} = H_{k_{DH}}(N_{AD1}\|N_{AD2}\|''1'')$ and a MAC key as $K_{a}= H_{k_{DH}}(N_{AD1} \|N_{AD2}\|''2'')$. We can further generate (session) keys in a similar manner for data stream-specific virtual links\footnote{Virtual Links (VLs): Each communication relationship in an aircraft network is represented as a VL. In our proposal we assume that a pair of communication parties would have two uni-directional VLs and each VL will have its own session key.} (VLs) for managing the communication between different aircraft sub-systems. 

Subsequently, the TPM generates a state validation message signed by the TPM AIK key represented in the protocol as ``$Sign_{TPM_{AD2}}(AD2-Validation)$''. AD2 will also request AD1 to provide assurance of its current state.

On receipt of this message, AD1 will first generate the session keys. AD1 will then verify AD2's signature and validation proof generated by the TPM of AD2. As the signature key belongs to the TPM of AD2, an attacker cannot masquerade this signature. By verifying the signature, AD1 can ascertain the current state (PCR value) is measured by the TPM of AD2. Now AD1 can verify whether the PCR value represents a trusted and secure state or not. Since our protocol pre-setup AD1 would have the PCR value of a trusted and secure state of AD2.

Furthermore, AD1 will check the values of Diffie-Hellman exponentials (\emph{i.e.} $g^{{r}_{AD1}}$ and $g^{{r}_{AD2}}$) and of the generated random numbers to avoid main-in-the-middle and replay attacks.

\paragraph{\bf Message 3}
\label{sec:Message3}
AD1 will then generate a message similar to message 2, a signature by AD1 and trust validation proof generated by its TPM.  

On receipt of the message, AD2 will verify the trust validation proof and generate keys. Furthermore, AD2 will also check the values of the Diffie-Hellman exponentials and of the generated random numbers to avoid man-in-the-middle and replay attacks.

\subsection{Post-Protocol Process}
\label{sec:Post_Protocl_Process}
The shared material generated from the Diffie-Hellman exponential can be used to generate more keys than just the session encryption and MAC keys of the protocol. If this is not desirable then session encryption and MAC keys can be saved as master session keys. Individual VL keys can then be generated from these session keys. Based on the security policies related to the VLs, whether they require only confidentiality or integrity or both, these two master session keys can be used to generate VL specific encryption and MAC keys. 

\subsection{Protocol Resumption}
\label{sec:Protocol_Resumption}
As discussed in \cite{RNAkram2015}, secure channel protocols only run when an aircraft is stationary on the ground, with proofs that the aircraft is not in flight based on geo-location, proximity to airport, weight on wheels, etc. The proposed protocol would run before each flight and master session keys are only valid for a single flight. The protocol should not be executed during the flight. Therefore, if a device has to reset due to some unforeseeable situation, a safety procedure to resume the secure channel and all of the associated VL keys - without running the protocol - must exist. For this purpose, each individual device will save the master session keys in its persistent storage and will have a standard algorithm to generate the keys for each of the VLs. If the master session keys are lost, then, during that particular flight, the device would be out of communication. To avoid this, the master session keys should be stored on two different memories (each aircraft device has at least two separate storage media, so as to provide this dual storage redundancy).

\section{Protocol Evaluation}
\label{sec:Protocol_Evaluation}
In this section, we first discuss the information analysis of the protocols, and then compare different protocols with our proposal based on the comparison criteria defined above. Finally, we provide some implementation results and a formal analysis using CasperFDR. 

\subsection{Brief Information Analysis}
\label{sec:Brief_Information_Analysis}
Throughout this section, we refer to the protocol comparison criteria of section \ref{sec:Comparison_Criteria} by their respective numbers as listed in the same section.

During the proposed protocol, in messages 2 and 3 the communicating entities authenticate each other, which satisfies G1. Similarly, for G2, all communicating entities have exchanged cryptographic certificates to facilitate an authentication and trust validation proof (generated and signed by the TPM) before the aircraft devices are deployed (pre-deployment configuration). 

The proposed protocol satisfies requirements G3, G4, G5 and G12 by first requiring AD1 and AD2 to generate the Diffie-Hellman exponential; thus computational cost is equal on both sides. Similarly, exponential generation also assures that both devices will have equal input to the key generation. Messages 2 and 3 are encrypted used the keys generated during the protocol, thus providing mutual key confirmation (satisfying G6).

In the proposed protocol, session keys generated in one session have no link with the session keys generated in other sessions, even when the session is established between the same devices. This enables the protocol to provide resilience against known-key security (G7). This unlinkability of session keys is based on the fact that each entity not only generates a new Diffie-Hellman exponential but also a random number, both of which are used during the protocol for key generation. Therefore, even if an adversary ``$\mathcal{A}$'' finds out about the exponential and random numbers of a particular session, it will not enable him to generate past or future session keys. 

Furthermore, to provide unknown key share resilience (G8), the proposed protocol includes the Diffie-Hellman exponentials along with generated random numbers and each communicating entity then signs them. Therefore, the receiving entity can then ascertain the identity of the entity with which it has shared the key. 

The protocol can be considered to be a KCI-resilient (G9) protocol, as protection against the KCI is based on the digital signatures. In addition, the cryptographic certificates of each signature key also include its association with a particular device. Therefore, if $\mathcal{A}$ has knowledge of the signature key of a device, it can only masquerade this particular device to other devices but not others to it. %Another point to note is that during the STCP, all signed messages and certificates are encrypted using the session key. This facilitates the STCP in meeting the requirements eight, and nine; as an adversary cannot substitute the certificate or signature.

\begin{table*}
\caption{Protocol comparison on the basis of the stated goals (see section \ref{sec:Comparison_Criteria}.)} 
\label{tab:ProtocolComparisonOnTheBasiesOfStatedGoals}

\begin{center}
			\begin{tabular}{!{\vrule width 1pt}@{ }r@{ }l!{\vrule width 0.75pt}c|c|c|c|c|c|c|c|c|c|c|c|c!{\vrule width 1pt}}
\noalign{\hrule height 1pt}
 			\multicolumn{2}{!{\vrule width 1pt}l!{\vrule width 0.75pt}}{\multirow{2}{*}{\bf Goals}} &  \multicolumn{13}{c!{\vrule width 1pt}}{\bf Protocols }  \\ \cline{3-15}
    && STS   & AD    & ASPeCT& JFK   & T2LS  & SCP81 & MM   & SM    & Asymmetric TKDF & P-STCP & SSH   & SSL   & Proposed Protocol \\ \noalign{\hrule height 0.75pt}
 G1.&& $*$   & $*$   & $*$   & $*$   & $*$   & $*$   & $-*$ & $-*$  & $*$   & $*$ & $(*)$ & $*$   & $*$   \\ \hline 
 G2.&& $*$   & $*$   & $*$   & $*$   & $*$   & $*$   & $*$  & $-*$  & $*$   & $*$ & $*$   & $*$   & $*$   \\ \hline  
 G3.&& $*$   & $*$   & $*$   & $*$   & $*$   & $*$   & $*$	& $-*$  & $-*$  & $*$ & $*$   & $*$   & $*$   \\ \hline  
 G4.&& $*$   & $*$   & $*$   & $*$   & $(*)$ & $*$   &      &		& $-*$  & $*$ & $(*)$ & $(*)$ & $*$   \\ \hline  
 G5.&& $*$   & $*$   & $*$   & $*$ 	 & $*$ 	 & $*$   & $*$  & $-*$  & $*$   & $*$ & $*$   & $*$   & $*$   \\ \hline  
 G6.&& $*$   &       & $*$   & $*$ 	 &  	 &		 & $*$  & $-*$	& $*$   & $*$ & $*$   & $*$   & $*$   \\ \hline  
 G7.&& $*$	 & $*$   & $*$   & $*$ 	 & $*$ 	 & $*$   & $*$  & 	    & $*$   & $*$ & $*$   & $*$   & $*$   \\ \hline  
 G8.&& $*$   & $*$   & $*$   & $*$ 	 & $*$ 	 & $*$   & $*$  & $-*$	& $-*$  & $*$ & $*$   & $*$   & $*$   \\ \hline 
 G9.&& $*$   & $*$   & $*$   & $*$ 	 & $*$   & $*$   & $*$  & $*$  	& $*$   & $*$ & $*$   & $*$   & $*$   \\ \hline
G10.&& $*$   &       & $*$   & $*$ 	 & $*$ 	 & $*$   &      &      	& $*$   & $*$ & $*$   & $*$   & $*$   \\ \hline 
G11.&& $*$	 &       &       & $*$ 	 & $*$   & $*$   & $*$  & $*$   & $*$   & $*$ & $*$   & $*$   & $*$   \\ \hline 
G12.&& $(*)$ & $(*)$ & $(*)$ & $(*)$ & $(*)$ & $(*)$ &      &       & $*$   & $*$ & $*$   & $*$   & $*$   \\ \hline 
G13.&&	     &       & $(*)$ & $(*)$ & $*$ 	 & $-*$  &      &       &       & $*$ & $(*)$ & $(*)$ & $*$   \\ \hline 
G14.&&       &       &       & $*$ 	 & $(*)$ &		 &      &       &       & $*$ & $(*)$ & $(*)$ & $*$   \\ \hline 
G15.&& $(*)$ &       & $*$   & $*$ 	 & $(*)$ &	     &      &       & $(*)$ & $*$ & $(*)$ & $(*)$ & $*$   \\ \hline 
		\end{tabular}
\end{center}		
	{\bf Note: }{$*$ means that the protocol meets the stated goal, $(*)$ shows that the protocol can be modified to satisfy the requirement, and $-*$ means that the protocol (implicitly) meets the requirement not because of the protocol messages but because of the prior relationship between the communicating entities.}

		%\vspace{-25pt}
\end{table*}

The proposed protocol also meets the requirement for perfect forward secrecy (G10) by making the key generation process independent of any long-term keys. The session keys are generated using fresh values of Diffie-Hellman exponentials and random numbers, regardless of the long term keys: they are signature keys. Therefore, even if eventually $\mathcal{A}$ finds out the signature key of any entity it will not enable him to determine past session keys. This independence of long term secrets from the session key generation process also enables the protocol to satisfy G7. 

Communicating entities in the STCP share signed messages with each other that include the session information, thus providing mutual non-repudiation (G11). G14 is ensured by the inclusion in the protocol of the session cookie, which provides a limited protection against DoS, and by the fact that individual devices have pre-configurations of communication partners which enable them to drop a connection if an entity trying to connect with them is not able to authenticate. 

To satisfy G15, the device identities are basically a random string that should not have any link with the function of the device. This would hinder an attacker from eavesdropping a protocol run to determine which aircraft device is communicating on the wireless channel. 

Finally, TPMs on all communicating devices provide trust validation proof in the form of PCR values signed by the TPM AIK. This provides mutual validation of the trust between communicating devices, confirming that the other device is functioning in a secure and reliable state (G13).

\subsection{Revisiting the Requirements and Goals}
\label{sec:Revisiting_the_Requirements_and_Goals}

Table \ref{tab:ProtocolComparisonOnTheBasiesOfStatedGoals} provides a comparison between the listed protocols in section \ref{sec:RelatedWorkonSecureChannelProtocols} with the proposed protocol in terms of the required goals (see section \ref{sec:Comparison_Criteria}). 

As shown in Table \ref{tab:ProtocolComparisonOnTheBasiesOfStatedGoals}, the STS protocol meets the first eleven goals. The main issue with the STS protocol is that it does not provide adequate protection against partial chosen key attacks (G12) and privacy protection (G15). The remaining goals are not met by the STS because of the design architecture and deployment environment, which did not require these goals. Similarly, the AD protocol does not meet G6, G10 and G13-G15.

The ASPeCT and JFK protocols meet a large set of goals. Both of these protocols can be easily modified to provide trust assurance (requiring additional signatures). Both of these protocols are vulnerable to partial chosen key attacks. However, in Table \ref{tab:ProtocolComparisonOnTheBasiesOfStatedGoals} we opt for the possibility that the ASPeCT and JFK protocols can be modified to meet this goal because in an AWN all communicating devices may be of the same computation power and have a strong offline pre-deployment relationship.

The T2LS protocol meets the trust assurance goal by default. However, for the remaining goals it has the same results as the SSL protocol. A point in favour of the SCP81, MM, and SM protocols is that they were designed for the smart card industry where there is a strong and centralised organisational model. Most of these protocols, to some extent, have a similar architecture, in which a server generates the key and then communicates that key to the client. There is no non-repudiation as they do not use signatures in the protocol run.

Both SSH and SSL meet a large set of requirements and also have the potential to be extended to the additional requirements. However, to provide a flexible, backward compatible and universally acceptable architecture these protocols have too many optional parameters. Such flexibility is one of the main causes of most of the issues that these protocols have been plagued with in the last couple of years, heartbleed being the most infamous vulnerability.  

Asymmetric TKDF (Trusted Key Distribution Frameworks) does not satisfy a number of requirements. In contrast, P-STCP satisfies most of the requirements listed in the table. The only difference between the P-STCP and the proposed protocol (except for the message structure) is the number of rounds to successfully complete a protocols run. P-SCTP has four messages (2-round protocol) and the proposed protocol uses 3 messages (1.5-round protocol). 

As apparent from the table \ref{tab:ProtocolComparisonOnTheBasiesOfStatedGoals}, the proposed protocol satisfies all goals that were described in section \ref{sec:Comparison_Criteria}.

\subsection{Practical Implementation}
\label{sec:Practical_Implementation}
In our AWN test-bed each node is a Raspberry Pi model B supplied with a Wi-Fi USB dongle TL-WN722N by TP-LINK.
In all the measurements we made, the nodes were configured in ad-hoc mode.

For all the selected protocols, in our evaluation implementations, we setup two neighboring nodes to establish a secure channel. This provides a performance measurement of the protocols between individual communicating pairs. 
However for the TKDF, a key distribution server is also required and a third node in the ad-hoc network plays this role.

\begin{figure}[htbp]
	\centering
		\includegraphics[width=0.90\columnwidth]{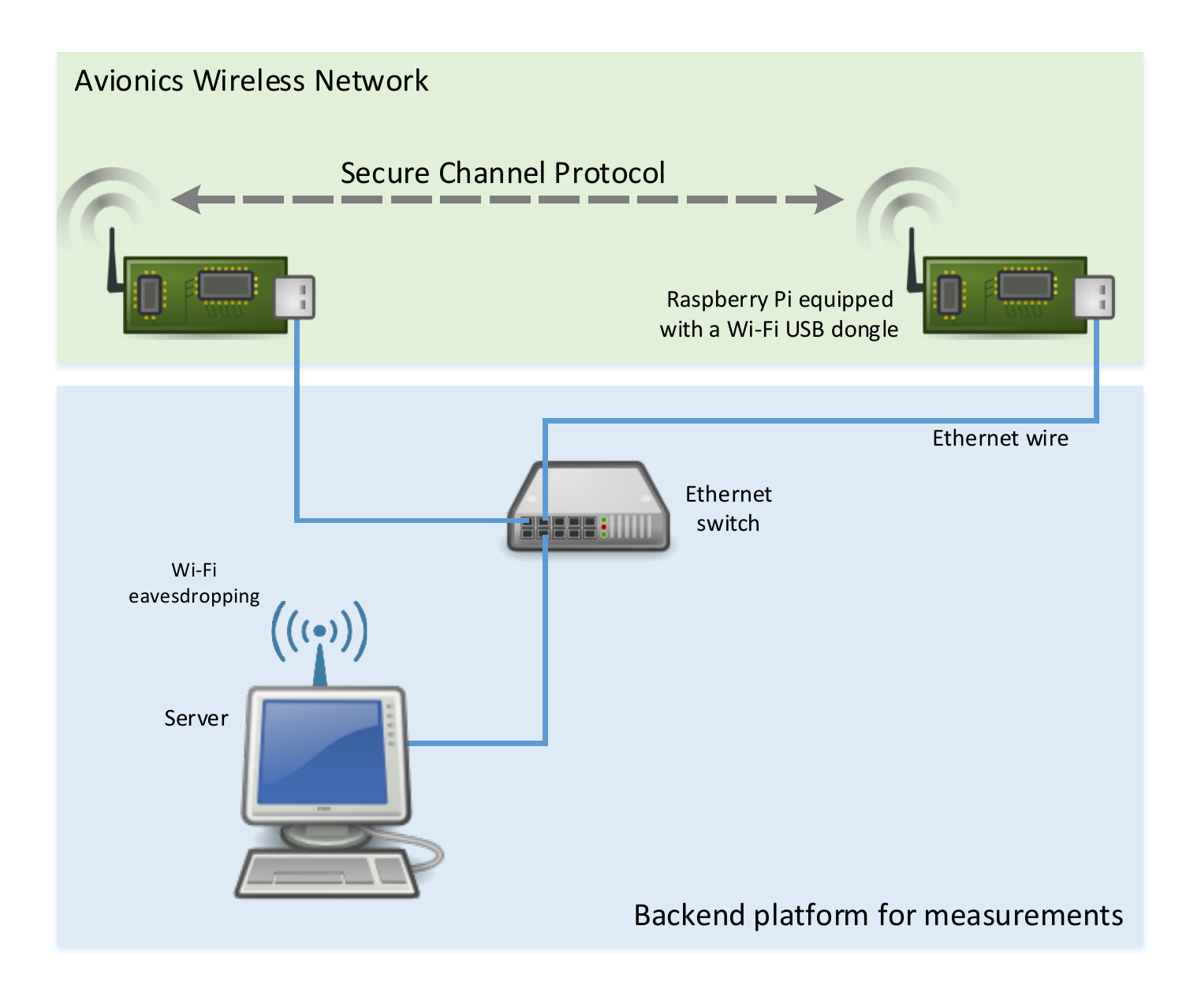}
	\caption{AWN test-bed}
	\label{fig:AWN test-bed}
\end{figure}

In our AWN test-bed, each node is connected to a backend server by means of an Ethernet connection.
This server controls the nodes so as to prepare them for the target scenario and is also in charge of collecting the measurements.
Effective measurement can be done internally on the node initiating the secure channel, called a client, and/or it can be done at the level of the network data exchanged between the nodes of the AWN and captured with a Wi-Fi card on the backend server set in monitor mode.

The performance comparison is provided in Table \ref{tab:Performance Measures}, comparing a subset of protocols from table \ref{tab:ProtocolComparisonOnTheBasiesOfStatedGoals} and proposed protocol performance in the developed test-bed environment.

\begin{table} [h]
	\begin{center}
\caption{Protocol performance measures (milliseconds)}
	\label{tab:Performance Measures}
\resizebox{0.95\columnwidth}{!}{%
	\begin{tabular}{|c|c|c|c|c|c|}
		 
		 \hline
         {\bf SSL} & {\bf SSH} & {\bf Asymmetric TKDF}& {\bf Proposed Protocol}\\ \hline
	     1310.93   & 911.21    & 14447.63 & 4582.44\\ \hline
		\end{tabular}
	}\vspace{5pt}	
				{\bf Note:}{ Above-mentioned measurement values for SSL, SSH and Asymmetric TKDF are from \cite{RNAkram2016g}}.
                
	\end{center}
\end{table}

In our Python implementation of the proposed protocol, the TPM was emulated by the Raspberry Pi.
Key sizes used for our proposed protocol were 2048 bits MODP group for the Diffie-Hellman key generation, 2048 bits for RSA and 256 bits for symmetric encryption and MAC computation (AES).

The P-STCP protocol was implemented with smaller key sizes in \cite{Akram2012b}, resulting in 2998.71ms performance measurement. Use the key sizes from \cite{Akram2012b} in our implementation results the performance of the proposed protocol to be 1201.50ms. 

%The protocol P-STCP, as in Table \ref{tab:ProtocolComparisonOnTheBasiesOfStatedGoals} meets similar number of requirements as the proposed protocol. The P-STCP in \cite{Akram2012b} has a performance measurement of 2998.71ms. The key sizes used in P-STCP implementation \cite{Akram2012b} were smaller then the ones selected used in table \ref{tab:Performance Measures}. If we used the key sizes selected in \cite{Akram2012b} then the performance measurement for the proposed protocol would be 1200ms.  %however, the key sizes were smaller then out protocol's implementation presented in Table \ref{tab:Performance Measures}. %If we implement the proposed protocol with similar key sizes a P-SCTP then the performance results are 1200ms. 

\subsection{Protocol Verification by CasperFDR}
\label{sec:CasperFDRProVerif}
We selected the CasperFDR approach for formal analysis of the proposed protocol. The Casper compiler \cite{CasperFDR1998} takes input as a high-level description of the protocol, together with its security requirements along with the definition of an attacker and its capabilities. The compiler then translates the description into the process algebra of Communicating Sequential Processes (CSP) \cite{CSPBook1978}. The CSP description of the protocol can be machine-verified using the Failures-Divergence Refinement (FDR) model checker \cite{CSP_Approach2000}. The intruder's capability modelled in the Casper script (appendix \ref{app:CasperFDR Script}) for the proposed protocol is: 

\begin{itemize}
\item an intruder can masquerade any entity in the network,
\item an intruder can read the messages transmitted in the network, and 
\item an intruder cannot influence the internal process of an entity in the network.
\end{itemize}

The security specification for which CasperFDR evaluates the network is as shown below. The listed specifications are defined in the \#Specification section of appendix \ref{app:CasperFDR Script}: 

\begin{itemize}
\item the protocol run is fresh and both applications are alive,
\item the key generated by the entity A is known only to the entity B (A and B are communication partners/devices),
\item entities mutually authenticate each other and have mutual key assurance at the conclusion of the protocol,
\item long-term keys of communicating entities are not compromised, and 
\item an intruder is unable to deduce the identities from observing the protocol messages.
\end{itemize} 

The CasperFDR tool evaluated the protocol and did not find any feasible attack(s). The script is provided in appendix \ref{app:CasperFDR Script}.

\section{Conclusion and Future Research Directions}
\label{sec:Conclusion}

In this paper, we outlined the concept of the AWN and discussed why such a proposal requires a secure channel for communication. The data communicated over an AWN has a strong requirement for confidentiality and integrity. To satisfy this requirement, communicating devices should have some cryptographic secrets to provide confidentiality and integrity. To generate these cryptographic secrets, the devices run a secure channel protocol. In this paper, we proposed a secure channel protocol that not only provides mutual authentications and key sharing between the communicating entities but also provides assurance that each of the devices is in a secure and trusted state. We compared our proposed protocol with a list of selected protocols and experimental performance results were provided. Finally, we evaluated the protocol using CasperFDR, showing that our protocol is secure against a number of attacks.

In future work, we will explore the major issue of detecting and neutralising wireless jamming and DoS attackers, along with building a strong mitigating framework. In addition to the trusted boot, for robust and reliable security we need to look into secure execution on AWN nodes - especially investigating the inclusion of ARM TrustZone and Intel SGX technologies.

\section{Acknowledgments}

The authors from Royal Holloway University of London acknowledge the support of the UK's innovation agency, InnovateUK, and the contributions of the  SHAWN project partners. The authors from XLIM acknowledge the support of:
\begin{itemize}
\item the SFD (Security of Fleets of Drones) project funded by R\'egion Limousin;
\item the TRUSTED (TRUSted TEstbed for Drones) project funded by the CNRS INS2I institute through the call 2016 PEPS (``Projet Exploratoire Premier Soutien'') SISC (``S\'ecurit\'e Informatique et des Syst\`emes Cyberphysiques'');
\item the SUITED (Suited secUrIty TEstbed for Drones) and UNITED (United NetworkIng TEstbed for Drones) projects funded by the MIRES (Math\'ematiques et leurs Interactions, Images et information num\'erique, R\'eseaux et S\'ecurit\'e) CNRS research federation;
\end{itemize}
The authors from LaBRI acknowledge the support of:
\begin{itemize}
\item the TRUSTED (TRUSted TEstbed for Drones) project funded by the CNRS INS2I institute through the call 2016 PEPS (``Projet Exploratoire Premier Soutien'') SISC (``S\'ecurit\'e Informatique et des Syst\`emes Cyberphysiques'');
\item the SUITED-BX and UNITED-BX projects funded by LaBRI and its MUSe team.
\end{itemize}

\section*{Disclaimer}
The views and opinions expressed in this article are those of the authors and do not necessarily reflect the position of SHAWN project or any of organisations associated with this project. 

%\printbibliography
%\bibliographystyle{IEEEtran}
%\bibliography{main}
%\include{main.bbl}
%\printbibliography


\appendices
\section{CasperFDR Script}
\label{app:CasperFDR Script}

{\fontsize{9}{0} 
\begin{spverbatim}
#Free variables
datatype Field = Gen | Exp(Field, Num) unwinding 2
hkAD2, hkAD1, iMsg, rMsg, EnMaKey : Field
AD1, AD2, U: Agent
gAD1, gAD2: Num
nAD1, nAD2, AD1Val, AD2Val: Nonce
VKey: Agent->PublicKey
SKey: Agent->SecretKey
InverseKeys = (VKey, SKey), (EnMaKey, EnMaKey), (Gen, Gen), (Exp, Exp)
\end{spverbatim}

\begin{spverbatim}
#Protocol description
0.    -> AD2 : AD1 [AD1!=AD2] <iMsg := Exp(Gen,gAD2)>
1. AD2 -> AD1 : AD2, nAD2, iMsg%hkAD2 <EnMaKey := Exp(hkAD2, gAD1); rMsg := Exp(Gen,gAD1)>
2. AD1 -> AD2 : nAD1, rMsg%hkAD1 <EnMaKey := Exp(hkAD1, gAD2)>
3. AD2 -> AD1: nAD2, nAD1
4. AD1 -> AD2 : {{rMsg, U, nAD2}{SKey(U)}}{EnMaKey} [rMsg==hkAD2]
5. AD2 -> AD1 : {{iMsg,AD2, nAD1}{SKey(AD2)}}{EnMaKey} [iMsg==hkAD1]
6.AD1 -> AD2 : {{AD1OSHash, AD1, nAD2}{SKey(AD1)}}{EnMaKey}
\end{spverbatim}

\begin{spverbatim}
#Actual variables
ADev1, ADev2, ME: Agent
GAD1, GAD2, GMalicious: Num
NAD1, NAD2, AD1VAL, AD2VAL, NMalicious: Nonce
\end{spverbatim}

\begin{spverbatim}
#Processes
INITIATOR(AD2,AD1, U, AD2VAL, gAD2, nAD2)knows SKey(AD2), VKey
READ2ONDER(AD1,AD2, U, AD1VAL, gSC, nSC) knows SKey(U), SKey(SC), VKey
\end{spverbatim}

\begin{spverbatim}
#System
INITIATOR(ADev2, ADev1,ADev2Val, GAD2, NAD2)
READ2ONDER(ADev1, ADev2, ADev1Val, GAD1, NAD1)
\end{spverbatim}

\begin{spverbatim}
#Functions
symbolic VKey, SKey 
\end{spverbatim}

\begin{spverbatim}
#Intruder Information
Intruder = ME
IntruderKnowledge = {ADev2, ADev2, ME,
GMalicious, NMalicious, SKey(ME), VKey}
\end{spverbatim}

\begin{spverbatim}
#Specification
Aliveness(AD2, AD1)
Aliveness(AD1, AD2)
Agreement(AD2, AD1, [EnMaKey])
Secret(AD2, EnMaKey, [AD1])
Secret(AD1, U, [AD2])
\end{spverbatim}

\begin{spverbatim}
#Equivalences
forall x, y : Num . Exp(Exp(Gen, x), y) = Exp(Exp(Gen, y), x)
\end{spverbatim}
}

%}
% \vspace{0.5cm}
% \centering
% \emph{\large 35th Digital Avionics Systems Conference\\September 25--29, 2016}

%%%%%%%%%
% Uncomment the following two lines for having footnotes at the end of document
%\newpage
%\theendnotes
%%%%%%%%%

\end{document}